\documentclass[sigconf]{acmart}

\newcommand{\POST}[1]{\texttt{POST}}
\newcommand{\GET}[1]{\texttt{GET}}
\newcommand{\PATCH}[1]{\texttt{PATCH}}
\newcommand{\DELETE}[1]{\texttt{DELETE}}
\newcommand{\dquote}[1]{{``#1''}}

\usepackage{caption}
\usepackage{subcaption}
\usepackage[vskip=1em,font=itshape,leftmargin=2em,rightmargin=2em]{quoting}
\usepackage{amsmath}

\usepackage{cleveref}
\crefformat{section}{\S#2#1#3}
\usepackage{multirow}
\usepackage{array} 
\usepackage{makecell}

\AtBeginDocument{%
  \providecommand\BibTeX{{%
    \normalfont B\kern-0.5em{\scshape i\kern-0.25em b}\kern-0.8em\TeX}}}

\copyrightyear{2026}
\acmYear{2026}
\setcopyright{cc}
\setcctype{by}
\acmConference[CHI EA '26]{Extended Abstracts of the 2026 CHI Conference on Human Factors in Computing Systems}{April 13--17, 2026}{Barcelona, Spain}
\acmBooktitle{Extended Abstracts of the 2026 CHI Conference on Human Factors in Computing Systems (CHI EA '26), April 13--17, 2026, Barcelona, Spain}
\acmDOI{10.1145/3772363.3799319}
\acmISBN{979-8-4007-2281-3/2026/04}
\begin{document}

\title{Envisioning Alternative Futures for Delivery Work: Toward a Decentralized Network Built on the OpenCourier Protocol}
\author{Yuhan Liu}
\email{yl8744@princeton.edu}
\orcid{0000-0001-6852-6218}
\affiliation{%
  \institution{Princeton University}
  \city{Princeton}
  \state{New Jersey}
  \country{USA}
}

\author{Varun Nagaraj Rao}
\email{varunrao@princeton.edu}
\orcid{000-0002-4692-2196}
\affiliation{%
  \institution{Princeton University}
  \city{Princeton}
  \state{New Jersey}
  \country{USA}
}

\author{Sohyeon Hwang}
\email{sohyeon@princeton.edu}
\orcid{0000-0001-8415-7395}
\affiliation{%
  \institution{Princeton University}
  \city{Princeton}
  \state{New Jersey}
  \country{USA}
}

\author{Janet Vertesi}
\email{jvertesi@princeton.edu}
\orcid{0000-0003-4579-6252}
\affiliation{%
  \institution{Princeton University}
  \city{Princeton}
  \state{New Jersey}
  \country{USA}
}

\author{Andrés Monroy-Hernández}
\email{andresmh@princeton.edu}
\orcid{0000-0003-4889-9484}
\affiliation{%
  \institution{Princeton University}
  \city{Princeton}
  \state{New Jersey}
  \country{USA}
}



\begin{abstract}
In this vision paper, we outline a blueprint for a decentralized network for the delivery industry, powered by an open protocol. By presenting the network’s key components and layers, alongside hypothetical scenarios, we illustrate how the network and the protocol may function in practice. Through this decentralized approach, we aim to address three major issues that mark the current platform-based delivery economy: power imbalances between the platform and workers, information asymmetries caused by opaque decision-making, and value misalignments. Our goal is to provoke dialogue and inspire future work toward more equitable, transparent, and worker-centered futures in the delivery industry, the broader gig economy, and related domains.
\end{abstract}





\maketitle
\begin{figure}
    \centering
    \includegraphics[width=0.9\linewidth]{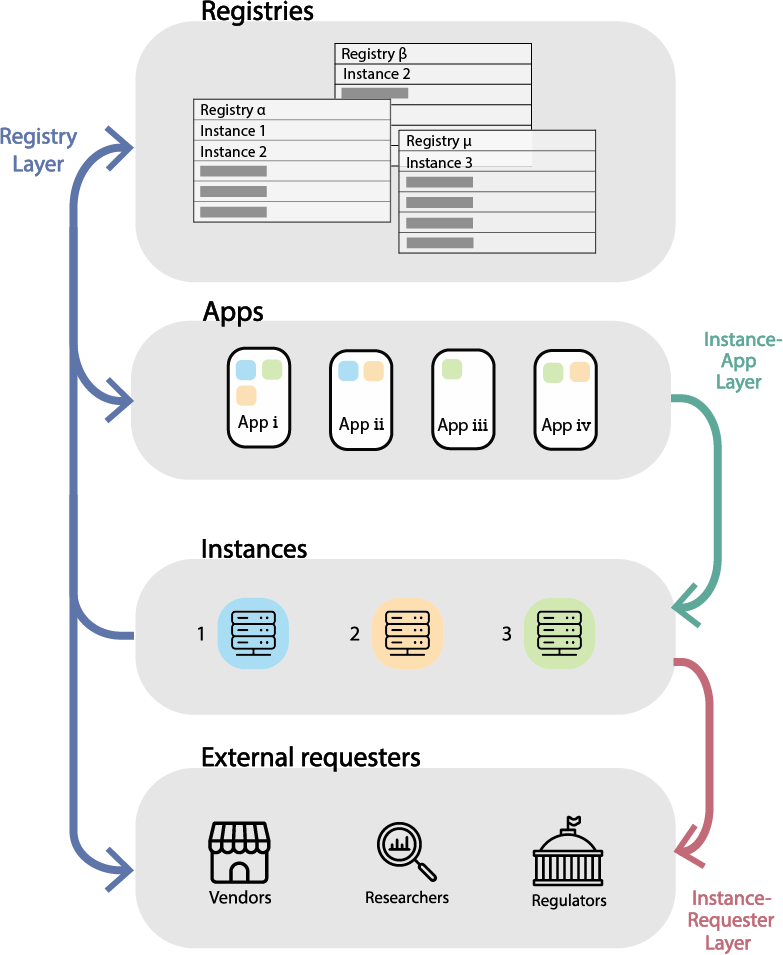}
    \caption{A Conceptual Overview of the Architecture for the Decentralized Delivery Network. See \cref{sec:architecture} for details.}
    \label{fig:teaser}
\end{figure}

\section{Introduction}
Gig work platforms like Uber and DoorDash have changed the landscape of work, offering people new ways to earn money and request convenient on-demand services~\cite{friedman2014workers, parigi2016gig, alkhatib2017examining}. In the United States, 36-38\% of the work-age population has engaged in gig work across diverse industries such as freelancing, transportation, food delivery, home services, and crowdwork~\cite{upwork2023freelance, garin2023evolution, cnn2023gig, deloitte_gig_economy_shared_services}. Although the flexibility in choosing when and how to work has motivated many individuals to join~\cite{ wood2019good, ens2018decent, rosenblat2016regional, hall2018analysis}, the centralized operational model --- where the platform sets the rules and uses algorithms to make key decisions like pricing and job assignment --- of the most widely used platforms, can concentrate and obscure decision-making in ways that undermine the well-being of workers \cite{nagarajrao2024navigating, bucher2021pacifying}. 

Industry practitioners and researchers have sought to mitigate these concerns. In recent years, \dquote{indie} food delivery platforms have emerged as alternatives that operate locally, intentionally hire local workers, and offer competitive compensation to support greater worker agency~\cite{liu2024mapping, dalal2023understanding}. The proliferation of these platforms across the United States provides evidence of the viability of decentralized, community-owned infrastructures~\cite{liu2024mapping, sultan2025comparative}. However, building more platforms requires technical expertise and engineering support. Indie platforms thus often rely on white-labeled software\footnote{White-label software refers to software developed by the vendor but rebranded and used by another organization as if it were their own. It is typically licensed through a subscription or per-transaction fee model. End users generally remain unaware of the original software provider.} (primarily from two vendors: DataDreamers\footnote{\texttt{https://datadreamers.com/}} and DeliverLogic\footnote{\texttt{https://www.deliverlogic.com/}}~\cite{liu2024mapping}) that can be expensive and offer only limited options for customization to fit local needs~\cite{liu2024mapping}. Additionally, indie platforms need consumers to know about and install an additional application, facing classic challenges of adoption and scale. These constraints are exacerbated by fragmentation: each indie platform must build and maintain a full stack and grow a user base largely in isolation. An open protocol can provide shared infrastructure so platforms can interoperate instead of rebuilding from scratch.

One promising path forward comes from recent moves to decentralize social media, which show that alternative models not only are possible but also can become widely used by leveraging open protocols \citep{masnick2019protocol}. Inspired by this push~\cite{twitter_mastodon_bloomberg, twitter_mastodon_mit_review}, we envision a decentralized network that shifts control away from monopolistic entities and toward a collectively governed network. This network is facilitated by an open {\em protocol}, which standardizes data formats and defines communication patterns to make interaction possible across a diverse ecosystem of delivery platforms, couriers, and service requesters. 

The goal of this paper is to outline a blueprint of the decentralized network of delivery platforms, enabled by an open protocol, which we provide details of in the Appendix. We also illustrate how the network and the protocol would work in action to improve the workers' well-being, focusing on three challenges that have persisted in the platform era. 




\section{Related Work: Issues in Mainstream Delivery Platforms}
Mainstream delivery platforms---the digital services that coordinate the logistics of transporting goods or people between couriers and service requesters--- have increasingly subjected workers to precarious conditions and limited protections~\cite{moore2021augmented, van2017platform, adermon2022gig, beerepoot2015competition}.\footnote{We term platforms like Uber, Doordash as ``mainstream'' platforms because they dominate the United States delivery landscape with a significant collective market share in recent years ~\cite{ahuja2021ordering}} Here, we review three key dynamics in these system that undermine workers' well-being and motivate our vision.

\subsection{Power Imbalance}
Workers typically have little influence over task allocation, compensation decisions, or avenues for recourse or feedback \cite{jarrahi2019algorithmic, jarrahi2020platformic, zwickWelcomeGigEconomy2018, rao2025fareshare}, while platform operators can control any of these. Platforms often leverage techniques such as algorithmic management (i.e., using algorithms to automate operations such as task dispatching, routing, payment, and performance monitoring) and gamification (i.e., using game design elements to incentivize workers toward earnings but possibly against their other interests), both of which can further decrease the agency and autonomy of workers. For example, prior research shows that automation in work contexts can contribute to physical and social isolation, weakening workers’ collective bargaining power~\cite{yao2021together, sahai2020workplace, li2023well, cieslik2022offline, zhu2021different}. This undermines workers' ability to choose and strategize around tasks~\cite{huang2023algorithmic, jarrahi2019algorithmic, lee2015working, rosenblat2016algorithmic}, particularly as tactics like surge pricing encourage workers to accept more tasks and work longer hours, even at a risk to their health~\cite{figueroa2021essential, simet2025gigtrap}.


\subsection{Information Asymmetry}
Platforms possess substantially more information than workers about the gig marketplace, the algorithms used to allocate tasks, and how fees are determined --- all contributing to pervasive information asymmetries between platforms and workers~\cite{rosenblat2016algorithmic}. Through mechanisms of ``soft control'' (e.g., surge pricing and incentive schemes that nudge workers toward specific behaviors) and opaque, black-box algorithmic decisions, platforms create structural imbalances that constrain workers' understanding of and influence over their work~\cite{viljoenDesignChoicesMechanism2021, tomassetti2016does, nagarajrao2024navigating}. For example, platforms can deactivate worker accounts without giving clear rationales \cite{schwartz2023deactivation}, limiting workers' ability to make informed decisions \cite{shapiroAutonomyControlStrategies2018a}. This is particularly concerning when workers are mistakenly or arbitrarily deactivated \cite{schwartz2023deactivation, rao2025fareshare}. 
Because information asymmetries stems from design choices that restrict the transparency and accessibility of key information \cite{bommasani2023foundation}, researchers have called for public data disclosure and platform transparency reports as a way to mitigate these harms \cite{nagarajrao2024navigating, calacci2026fairfare}.

\subsection{Value Misalignment}
Mainstream delivery platforms develop their digital infrastructures and policies internally to serve their interests as a company serving as an intermediary between workers and other stakeholders~\cite{jarrahi2021algorithmic, zuboff2015big}. Corporate priorities rather than worker needs or preferences thus structure the policies, processes, and infrastructures of platforms~\cite{jarrahi2021algorithmic, zuboff2015big}. As platforms seek venture capital funding, they are also accordingly incentivized to optimize algorithms for speed, order volume, and market growth ~\cite{DoorDashSharesSink, hochberg2012venture, zwick2022entrepreneurial}. This produces a persistent misalignment between platform values and workers' needs~\cite{dedema2024socio, jarrahi2021algorithmic, duggan2020algorithmic, weber2022new}, while creating little room for workers to assert their values in decision-making processes \cite{li2022bottom}.
\section{Architecture of the Decentralized Network}
\label{sec:architecture}
To respond to these issues, we propose an open protocol {\em OpenCourier} that establishes a set of data and communication standards between different types of stakeholders in the delivery industry. This protocol envisions a decentralized network of independent but interoperable delivery platforms, e.g., worker-owned or local indie platforms. 
Because we aim to open space for dialogue and reflection on these speculative possibilities rather than to introduce a concrete artifact, we provide a high-level overview of the network and its implications. Detailed protocol specifications are included in the Appendix for reference.

\subsection{Actors in the Decentralized Network}

Given the motivations of this work, the envisioned network centers delivery workers. We highlight three key actors in the network. 

A \textbf{Courier} is defined as: \textit{a worker who completes short-distance deliveries by receiving tasks through mobile apps and transporting goods using bikes, scooters, or cars.} 

Couriers operate as part of \textbf{Courier Instances}: \textit{independently operated hubs where couriers manage work (e.g., receiving tasks, updating status, and completing deliveries)}. 
The network assumes each courier belongs to one or more instances, which may take various organizational forms and scales. For example, an instance may consist of one courier running services alone, or a worker cooperative. Crucially, independently-owned instances in the network are interoperable: the protocol allows them to exchange information like courier delivery history and reputation data. Thus, workers can move across instances. Broadly, such consistency also means platform software designed by one instance could potentially be used any others.

Finally, \textbf{External Requesters} are defined as: \textit{any parties that interact with courier instances by initiating requests, including service requests (e.g., restaurants, retailers, or customers) and data requests (e.g., researchers, auditors, or government agencies).}

\subsection{Interactions in the Decentralized Network}
Activities in the network fall under three layers: the registry layer, the instance-app layer, and the instance-requester layer. Interactions in these layers are supported by the open protocol that standardizes the communication between different actors. 

\subsubsection{Registry Layer}
In traditional delivery markets, couriers often rely on word-of-mouth or an existing, centralized platform to find opportunities and contracts. In a decentralized network, however, no single platform has ubiquitous visibility or reputation, making discovery more challenging. Our vision envisions \textit{registries} that address this challenge by defining shared, queryable directories of active delivery service providers. Registries can be curated by anyone, including labor unions or local govenrment. This helps couriers find legitimate instances to join and find work through, as well as enables requesters to locate providers in a certain area.

\subsubsection{Instance-App Layer}
Couriers need to be able to receive and share information about their work with courier instances, allowing them to accept or reject tasks, update delivery statuses, and manage personal preferences (e.g., about what kinds of tasks they would like to do). As most delivery work at present is app-based, we anticipate couriers using mobile apps to do so. The Instance-App layer defines how courier-facing apps communicate with infrastructure used by courier instances. This layer is critical in allowing couriers to additionally have meaningful voice, with three key functions. {\em Order fulfillment} endpoints manage task and courier status transitions throughout the delivery process. {\em Preference input} endpoints allow couriers to specify their work preferences in detail. For example, couriers can indicate 
areas they'd like to work within, constraints like weight limits or allergens, or prioritizing surge-pricing orders. We envision courier instances allocating jobs based on these preferences, possibly through custom algorithms. {\em Community Notes} endpoints enable information sharing within the courier community through location-based notes, such as parking tips tied to specific locations for their peers to reference.

\subsubsection{Instance-Requester Layer}
The Instance–Requester layer (a) handles service requests and (b) enables data disclosure.

To dispatch service requests to couriers, courier instances and service requesters must exchange structured, machine-readable information about delivery tasks, including pickup and drop-off locations, timing constraints, and compensation terms. Courier instances may also negotiate with service requesters by rejecting/accepting the tasks with text messages. 

To comply with the regulation, courier instances must be able to export data for disclosure and auditing. Many U.S. cities, including Chicago and New York, now require anonymized data disclosure from rideshare companies to monitor pricing equity, enforce labor protections, and inform transportation policy~\cite{cityofchicago_tnp_trips_2018_2022, nyctlc_tripdata}. For small indie platforms, the administrative costs of doing so can be too high. 
We advocate for using open protocols to standardize data schemas, reducing the burden on courier instances. A shared schema also simplifies data auditing, not only for individual instances (i.e., internal reviews) but across the entire network. 
\section{Potential Benefits of the Decentralized Network}
Here, we provide sketches of hypothetical scenarios to articulate and discuss how our vision of a protocol-based, decentralized network addresses the challenges currently faced by centralized delivery platforms.

\subsection{Scenario: Grant Agency to Couriers}
Prior work describes worker agency in the gig economy as the power workers hold in their relationships with platforms, including the freedom to make decisions about their work and the ability to shape how they interact with platform systems~\cite{dedema2024socio, beigi2022steering, anwar2020hidden, woodside2021bottom}. Our envisioned network operationalizes this form of agency by enabling couriers to choose, combine, and move across instances that reflect their capacities, values, and goals. Because the system is built on an {\em open} protocol, workers can also choose whichever mobile application they prefer, for example, on their UI/UX preferences, as long as the app implements the protocol and integrates its endpoints. Consider the case of Bob, an experienced courier managing a chronic shoulder condition: through any mobile app that is compatible with the open protocol, Bob selects a worker-owned cooperative whose task-allocation algorithm allows him to flag his chronic condition and avoid physically intense tasks. However, he also joins a second instance where he doesn't share this information, so he can sign in and increase his workload and earnings when his shoulder condition is manageable. 
Instances can have different goals, scopes, purposes --- but through the shared infrastructure of the protocol, workers can move across them relatively seamlessly. As worker needs vary and evolve, being able to work more versus less in multiple instances may better satisfy their overall goals.

\subsection{Scenario: Improve Data Transparency}
Data transparency in the network is guaranteed by the nature of open protocols. As an illustrative example, Mallory, a labor-policy researcher, requests voluntary data contributions from multiple locally operated courier instances in New York, Chicago, and Seattle. Because the open protocol defines a standardized data disclosure and auditing endpoint, participating instances can consent and share anonymized delivery data (e.g., pay rates, working hours, and delivery distances) without building custom export pipelines for each request. The consistent data format allows Mallory’s team to conduct cross-city comparisons and identify disparities, such as couriers in a city earning less per mile while working longer hours for equivalent pay. These transparent, comparable findings support evidence-based policy recommendations that are later discussed with local workers and city officials and inform proposed local legislation. In this way, the network not only has reduced the technical burdens and barriers for instances to share data, but also makes it easy to aggregate data to make meaningful analyses that can inform policy and help audit work practices in this industry more broadly.


More generally, a protocol can mandate disclosure of key information by defining certain fields in endpoints as required and non-empty. For example, instance operators in the network must provide a rationale for courier deactivation because the deactivation endpoint in the underlying protocol makes the \dquote{reason} parameter mandatory. Here, we aim to counter the opacity often associated with platform operations.

\subsection{Scenario: Encourage Work-Centered Design and Technical Infrastructure Innovations}


The open nature of protocol underlying the network broadens who can build and shape the technical infrastructure that underpins delivery work. By making the specification open, anyone can develop and publish compatible tools or services, expanding the network and fostering more diverse, worker-aligned innovations. In one hypothetical scenario, Alice, a software developer, creates an open-source optimization algorithm that improves how delivery requests are batched and assigned during peak hours. After publishing the algorithm with an implementation of the open protocol, instances facing surge-time bottlenecks can adopt it directly without developing custom infrastructure because the algorithm interoperates with existing protocol-compliant systems. Operators in dense urban areas are able to handle higher-order volumes with fewer delays and failures. Although not all organizations have the technical expertise or resources to build better algorithms or interfaces, the protocol follows open source software principles to catalyze collaboration and innovation from a broader range of potential contributors. Additionally, instead of having to rely on one closed system, workers can opt in or seek tools that meet their values and needs. Broadly, in order to retain workers (who can move across instances), operators of courier instances are incentivized to adopt tools that better align systems with couriers' needs and well-being, creating a worker-centered feedback loop. 


\section{Future Work}
While we present a decentralized, protocol-based approach as an alternative future for delivery work, future work guiding implementation and adoption are crucial to make this future viable. Here we highlight two key areas.

\subsection{Designing a Governance Model}
A decentralized network introduces unpredictable dynamics of competition and collaboration among platforms. We believe the governance model of such a decentralized network would require exploration in the following perspectives: 1) \textit{How do instances manage themselves?} E.g., Instances need to decide how to protect individual couriers from harm, as well as how to deal with couriers who break collective rules. They must also set processes for setting, enforcing, and revising any such norms. 2) \textit{How do instances manage interactions with one another?} E.g., Instances may be geographically adjacent and even overlap in workers, but have different pay policies that lead to conflict and unhealthy competition. 3)\textit{How does the network manage its registry of courier collectives?} E.g., the network needs to set rules and processes to safeguard against instances providing illegal or harmful services and that couriers operating in bad faith are not simply hopping from instance to instance.

\subsection{Community Adoption and Adaptation of the Protocol}
A key challenge for any network is attracting a sufficient number of adopters. Many indie platforms already operate with established infrastructures for order management, task assignment, and payment processing. Joining the decentralized network requires them to adapt or extend their systems to support the open protocol. One promising path is the development of dynamic adaptors that translate between existing endpoints in the white-label software and the protocol, avoiding the need for wholesale system replacement and lowering integration barriers. Such adaptors could also enable interoperability across multiple protocol-based networks in the future. Similar tools already exist in decentralized social media—for example, BridgyFed and Nipy-Bridge~\cite{barrett2025bridgyfed, nipy2024nipybridge}—which connect otherwise incompatible protocols. By reducing adoption friction and demonstrating practical interoperability, adapters can accelerate the growth of the decentralized network that supports shared governance and worker-centered design.

Broadly, the sustainability of this network also requires enough consumers moving to platforms that operate as part of it, away from or in addition to existing centralized platforms. Future work must also consider how consumer-facing systems might integrate into this ecosystem of delivery work.
\section{Conclusion}
This vision paper presents a blueprint for a decentralized sociotechnical network for delivery, enabled by an open protocol. We identified three key issues in mainstream delivery platforms---power imbalances, information asymmetries, and value misalignments---rooted in centralized control and opaque operations of existing systems. We then provide a high-level description of a decentralized network that connects independently operated courier instances through an open protocol that standardizes communication and interactions. We illustrate how the network can enhance worker agency, supports transparent data practices, and encourages worker-centered design innovations. Rather than proposing a finalized system, this work uses speculative, forward-looking design to provoke conversations about more equitable and democratic futures in delivery work. By foregrounding open protocols and decentralized governance, we aim to inspire new research and practical experimentation toward a fairer, more transparent, and worker-driven gig economy.

\begin{acks}
We acknowledge support from the Decentralization of Power Fund at Princeton University. We thank several collaborators who contributed to this project including Nikola Mitic, Eduardo Moreno, Astrit Zeqiri, Mike Perhats and Gleidson Gouveia, Kristoffer Selberg, Jessica-Ann Ereyi, Angela Tan, and Julia Ying.
\end{acks}

\bibliographystyle{ACM-Reference-Format}
\bibliography{reference}

\appendix
\section{Protocol Design}
Our protocol design mirrors principles of data portability and interoperability seen in early internet protocols (e.g., SMTP for email, HTTP for the web) 
that define open interfaces and enable anyone to build compatible implementations~\cite{livitckaia2023decentralised, wei2024exploring, Banday2010APS}. It is also inspired by the recent popularity of decentralized social media and e-commerce protocols that reshape governance models around technologies~\cite{PrincetonDSM2024, hwang2025trust, oshinowo2025seeing}.
\subsection{Registry Layer}
A registry, at minimum, provides a basic list of active instances with detailed information, as shown in Table \ref{tab:registry_info}, including the name of the platform, the geographic location the platform operates in, and brief rules and description of the platform for couriers to know the platform's value and policy better. This helps couriers find legitimate instances to join and find work through, as well as enables requesters to locate providers in a certain area. However, note that registering is not required for an instance, and couriers will still be able to join an instance given their direct link.

Most data the registry stores originates from the instance itself (e.g., name, rules, fees), and the instance is always the source of truth of this data. Furthermore, to incorporate transparency and accountability, this data is publicly available at the instance's \texttt{GET /metadata} endpoint. The registry then has the flexibility to select which instance-provided data to use and display, while also incorporating data it defines independently. In a sense, a registry acts as a directory of latest cached data of instances that have chosen to register in that registry.
\begin{table}[hbt!]
\small
\centering
\begin{tabular}{p{0.28\linewidth} p{0.64\linewidth}}
\toprule
\textbf{Field} & \textbf{Description} \\
\midrule
Instance Name & The name of the instance. \\
Domain Name & The domain name of the platform. \\
Logo & The visual identity of the instance, displayed in client or mobile applications. \\
Operation Region & The region or area the platform operates in, represented in \texttt{GeoJSON} format. \\
Rules URL & A link to the rules of the instance. \\
Description URL & A link to a short summary written in the language above describing the instance's mission, values, and policies to help couriers understand its operation. \\
Terms of Service URL & A link to the instance’s Terms of Service.\\
Privacy Policy URL & A link to the instance’s Privacy Policy.\\
User Count & The number of users in the instance. \\
Updated At & The time the instance last updated their details or config.\\
Created At & The time the instance registered to the registry.\\
Last Fetched At & The time when data was last fetched from the instance.\\
Status & The status of the instance: could be ``verified'', ``pending'', ``suspended'', etc.\\
\bottomrule
\end{tabular}
\caption{Metadata of Each Courier Instance in a Registry}
\label{tab:registry_info}
\end{table}
In the network, we envision the possibility of multiple registries. 
A registry can be hosted in multiple ways: for example, as a hard-coded list embedded in courier mobile apps, on a blockchain to enable free, identifiable, and low-barrier registration for each instance, or by a trusted third-party entity that validates and maintains a list of legitimate businesses or courier collectives. When using the courier applications, couriers can either directly join an instance via their link, or browse registries for nearby or popular instances. The protocol standardizes the minimum data a registry contains to make this straightforward. However, the specific decisions for what data the registry will use, how information from a registry is displayed and filtered, and how often the data is updated based on the newest data from the instance, depends on each specific registry and the mobile app implementing the protocol. 

\subsection{Instance-App Layer} 
To carry out their work, couriers need to be able to receive and share information with courier instances, allowing them to accept or reject tasks, update delivery statuses, and manage personal preferences (e.g., about what kinds of tasks they would like to do). The Instance-App layer defines how courier-facing apps communicate with courier instances through a standardized set of APIs. It is the heart of the open protocol and includes three key distinct sets of endpoints: order-fulfillment, preference-input, and community-note. Here we'll provide the overview and introduce key features in each component. 

\subsubsection{Order-fulfillment Endpoints}
As shown in Table \ref{tab:order-fulfillment}, the order-fulfillment endpoints manage task and courier status transitions throughout the delivery process. Order statuses include: \textit{dispatched, accepted, rejected, canceled, picked up, delivered}, while courier statuses include: \textit{online, offline, last-call, on the way, arrived at pickup, arrived at dropoff}. Status updates are handled via \POST{} and \PATCH{} requests, and the latest status can be queried using \GET{}. The order-fulfillment endpoints function very similarly to what the existing delivery APIs do to manage the deliveries and can be easily adapted from the existing APIs. In the open protocol, we aim to standardize the endpoint definition that can facilitate interoperability within the network. 

\begin{table*}[h]
\small
\centering
\begin{tabular}{p{0.58\linewidth} p{0.38\linewidth}}
\toprule
\textbf{Endpoint} & \textbf{Function} \\
\midrule
\texttt{GET /api/admin/v1/deliveries/\{deliveryId\}} & Get details of a delivery. (admin login required)\\
\texttt{GET /api/courier/v1/deliveries/new} & List my new deliveries. (courier login required)\\
\texttt{GET /api/courier/v1/deliveries/in-progress} & List my in-progress details (courier login required).\\
\texttt{GET /api/courier/v1/deliveries/done} & List my finished deliveries. (courier login required)\\
\texttt{POST /api/courier/v1/deliveries/\{deliveryId\}/accept} & Accept a delivery. \\
\texttt{POST /api/courier/v1/deliveries/\{deliveryId\}/reject} & Reject a delivery. \\
\texttt{PATCH /api/courier/v1/deliveries/\{deliveryId\}/cancel} & Cancel a delivery. \\
\texttt{POST /api/courier/v1/deliveries/\{deliveryId\}/mark-as-dispatched} & Mark a delivery as dispatched. \\
\texttt{POST /api/courier/v1/deliveries/\{deliveryId\}/arrived-at-pickup} & Indicate courier arrival at pickup location. \\
\texttt{POST /api/courier/v1/deliveries/\{deliveryId\}/mark-as-picked-up} & Mark the item as picked up. \\
\texttt{POST /api/courier/v1/deliveries/\{deliveryId\}/mark-as-on-the-way} & Mark the courier as en route to dropoff. \\
\texttt{POST /api/courier/v1/deliveries/\{deliveryId\}/arrived-at-dropoff} & Indicate courier arrival at dropoff location. \\
\texttt{POST /api/courier/v1/deliveries/\{deliveryId\}/mark-as-delivered} & Mark the item as delivered. \\
\texttt{PATCH /api/courier/v1/deliveries/\{deliveryId\}/report-issue} & Report an issue with the delivery.\\
\bottomrule
\end{tabular}
\caption{Order-fulfillment Endpoints in the Open Protocol}
\label{tab:order-fulfillment}
\end{table*}
\subsubsection{Preference-Input Endpoints}
We designed a \textit{courier-setting} endpoint that allows couriers to specify their work preferences in detail. For example, couriers can indicate the types of merchants they prefer, aligning with their values or supporting specific communities (e.g., black-owned businesses). Couriers can also specify their preferences according to their strategies for order acceptance/rejection, such as favoring orders under a certain weight (e.g., below 15 lbs) or prioritizing tasks with surge pricing during order matching through the input interface on the mobile app. Parameter names and a corresponding example is shown in Table \ref{tab:preference-input}. Couriers preferences can be retrieved through \GET{}, and updates are made through through \PATCH{}. Note that we only define the endpoints for couriers to input their preferences to an instance; how individual instances incorporate these preferences and manage daily operations is a matter of their organizational practice and falls outside the scope of the protocol's design. The current preference options in the endpoint are designed based on the input of our industrial collaborator. We welcome feedback and input from practitioners and workers for more options to enrich the endpoints. 
\begin{table}[h]
\small
\centering
\begin{tabular}{
  >{\raggedright\arraybackslash}p{0.20\linewidth}  
  >{\raggedright\arraybackslash}p{0.16\linewidth}  
  >{\raggedright\arraybackslash}p{0.58\linewidth}  
}
\toprule
\textbf{Field} & \textbf{Type} & \textbf{Example Value} \\
\midrule
\texttt{deliveryPolygon} & \texttt{GEOJSON} & \makecell[t{{l}}]{%
\texttt{\{"type": "Polygon",}\\
\texttt{"coordinates": [[}\\
\texttt{[-74.6675, 40.3520],}\\
\texttt{[-74.6565, 40.3520],}\\
\texttt{[-74.6565, 40.3435],}\\
\texttt{[-74.6675, 40.3435],}\\
\texttt{[-74.6675, 40.3520]]]\}}%
} \\
\texttt{vehicleType} & \texttt{String} & \texttt{"BICYCLE"} \\
\texttt{preferredAreas} & \texttt{String[]} & \texttt{["Downtown Princeton", "Princeton Junction"]} \\
\texttt{shiftAvailability} & \texttt{Json} & \texttt{\{"monday": ["09:00-13:00"], "friday": ["17:00-21:00"]\}} \\
\texttt{deliveryPreferences} & \texttt{String[]} & \texttt{["small order", "medium order"]} \\
\texttt{foodPreferences} & \texttt{String[]} & \texttt{["vegan"]} \\
\texttt{earningGoals} & \texttt{Json} & \texttt{\{"maximize": "per delivery rate"\}} \\
\texttt{deliverySpeed} & \texttt{String} & \texttt{"REGULAR"} \\
\texttt{restaurantTypes} & \texttt{String[]} & \texttt{["black-owned business"]} \\
\texttt{cuisineTypes} & \texttt{String[]} & \texttt{["halal", "vegan"]} \\
\texttt{dietaryRestrictions} & \texttt{String[]} & \texttt{["NONE"]} \\
\bottomrule
\end{tabular}
\caption{Example Schema of Courier Preference Input}
\label{tab:preference-input}
\end{table}

\subsubsection{Community-Note Endpoints}
To enable information sharing within the courier community, the protocol includes endpoints for location-based notes. Couriers can leave notes, such as parking tips tied to specific locations for their peers to reference. They can also react to specific notes with emojis to confirm its validity. The community notes feature is inspired by information-sharing in the gig worker local online forums and offline gathering camps~\cite{yao2021together, qadri2022drivers}. Corresponding endpoints are shown in Table \ref{tab:location-notes}.

\begin{table}[t]
\small
\centering
\begin{tabular}{
  >{\raggedright\arraybackslash}p{0.60\linewidth}
  >{\raggedright\arraybackslash}p{0.34\linewidth}
}
\toprule
\textbf{Endpoint} & \textbf{Description} \\
\midrule
\texttt{POST /api/courier/v1/location-notes} & Create a note. \\
\texttt{GET /api/courier/v1/location-notes} & List all my notes. (courier login required) \\
\texttt{PATCH /api/courier/v1/location-notes/\{locationNoteId\}} & Update a note. \\
\footnotesize{\texttt{GET /api/courier/v1/location-notes/\{locationNoteId\}}} & Get details of a note. \\
\texttt{DELETE /api/courier/v1/location-notes/\{locationNoteId\}} & Delete a note. \\
\texttt{POST /api/courier/v1/location-notes/\{locationNoteId\}/react} & Add reaction to a note. \\
\bottomrule
\end{tabular}
\caption{Location Notes Endpoints in the open protocol}
\label{tab:location-notes}
\end{table}

\subsection{Instance-Requester Layer}
In order to be distributed among couriers, courier instances and service requesters must communicate about delivery tasks—and their associated details such as pickup and dropoff locations. Other external actors, such as regulators or researchers, also benefit from standardized access to work data (e.g., to improve transparency). The Instance-Requester layer defines these interactions, providing two main categories of endpoints: those for handling service requests and those for data disclosure and auditing.

\label{section:instance-eco}
\subsubsection{Courier Instance - Service Requester Interaction}
Interactions between instances and service requesters primarily revolve around order negotiation. Courier instances are dedicated to delivery-related tasks, while requesters manage order placement from customers. Within the open protocol, requesters initiate a quote that specifies task details such as pickup and drop-off locations, delivery deadline, and proposed compensation, as shown in Table \ref{tab:delivery-quote-schema}. Courier instances may then respond by accepting, rejecting, or providing counteroffers through an open text field, allowing for several rounds of negotiation. Meanwhile, the registry list enables requesters to query and broadcast quotes across multiple instances to compare service options. Once negotiation concludes, the task is finalized and assigned to a single courier instance.
\begin{table}[t]
\small
\centering
\begin{tabular}{
  >{\raggedright\arraybackslash}p{0.28\linewidth}
  >{\raggedright\arraybackslash}p{0.65\linewidth}
}
\toprule
\textbf{Field} & \textbf{Description} \\
\midrule
\texttt{quote} & Estimated delivery quote. \\
\texttt{quoteRangeFrom} & Lower bounds of quote range. \\
\texttt{quoteRangeTo} &  Upper bounds of quote range. \\
\texttt{feePercentage} & Commission fee requester takes. \\
\texttt{currency} & Currency of the task.\\
\texttt{duration} & Estimated delivery duration in minutes. \\
\texttt{distance} & Delivery distance. \\
\texttt{distanceUnit} & Unit of distance, e.g., \texttt{MILES}. \\
\texttt{pickupPhoneNumber} & Phone number at pickup location; optional. \\
\texttt{pickupName} & Name of the contact at the pickup location. \\
\texttt{dropoffPhoneNumber} & Phone number at dropoff location. \\
\texttt{dropoffName} & Name of the contact at the dropoff location. \\
\texttt{expiresAt} & Time when the quote expires. \\
\texttt{pickupReadyAt} & Earliest time when courier can pick up. \\
\texttt{pickupDeadlineAt} & Latest time the order must be picked up. \\
\texttt{dropoffReadyAt} & Earliest time when courier can dropoff. \\ 
\texttt{dropoffEta} & Estimated time of arrival at dropoff. \\
\texttt{dropoffDeadlineAt} & Latest time the delivery must be completed. \\
\texttt{orderTotalValue} & Total order value.\\
\texttt{pickupLocation} & Pickup location. \\
\texttt{dropoffLocation} & Dropoff location. \\
\bottomrule
\end{tabular}
\caption{Schema of a Quote of a Delivery}
\label{tab:delivery-quote-schema}
\end{table}

\subsubsection{Data Disclosure and Auditing}
Major U.S. cities such as Chicago and New York City have mandated anonymized data disclosure from rideshare companies to support goals like monitoring pricing equity, enforcing labor protections, and informing transportation policy~\cite{cityofchicago_tnp_trips_2018_2022, nyctlc_tripdata}. In alignment with these practices, we introduced endpoints that allow CSV data export from courier instances, reducing information asymmetry between platform operators (i.e., instance admins) and couriers. These exports also enable instance admins to share data with third-party auditors and to build dashboards that surface insights such as average hourly earnings across all couriers. Beyond individual platforms, the protocol’s standardized data schema supports auditing at the network level. For example, a researcher might get data donations from a random sample of courier instances to measure average pay in different regions. While the current implementation provides only basic CSV dumps with authentication, future iterations may work towards advanced endpoints that grant privileged query access while embedding stronger privacy protections.

\section{Reference Implementation}
Here, we provide a basic reference implementation of the protocol as a proof-of-concept of the network. We describe a hypothetical instance registry, that is hard-coded into an app; a courier mobile app with the registry that couriers in an instance included on the registry can use; and a backend server application for instances, with an admin interface. 

\subsection{Registry}
\begin{figure}[hbt!]
    \centering
    \includegraphics[width=0.8\linewidth]{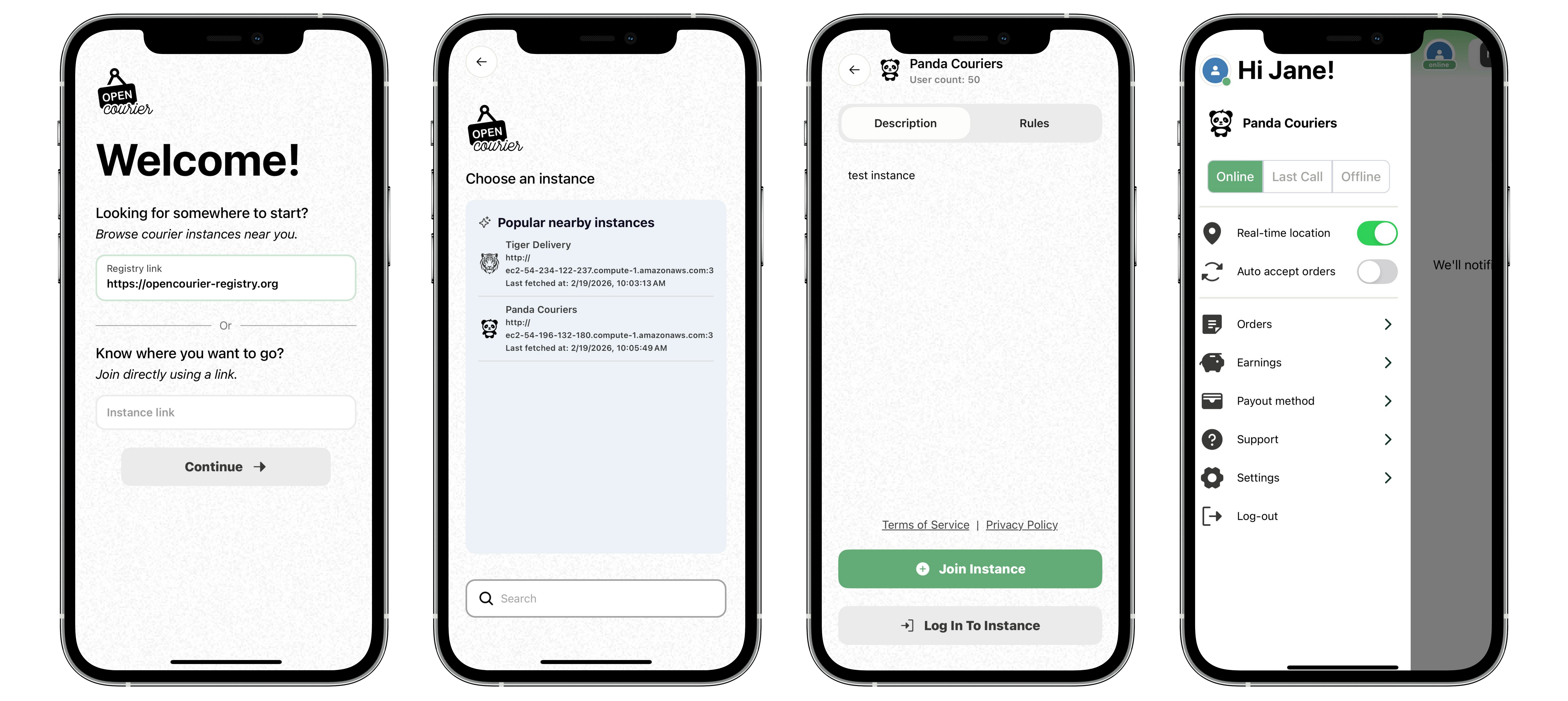}
    \caption{Screenshots of the onboarding pages showing the demo registry link, details of an instance, and how it displays instance name and icon in the side menu after logging in.}
    \label{fig:registry_mobile}
\end{figure}

We implemented a simple registry using a Node.js backend with Express and a PostgreSQL database with PostGIS extension enabled (for geospatial queries). This registry link is hard-coded within the mobile app to act as a demo, shown in Figure~\ref{fig:registry_mobile}. Administrators can register in this demo registry via the administrator interface (mentioned later). The mobile app uses the registry's API to display the instances sorted by distance to the courier, if their location is enabled.

Couriers also have the ability to directly join an instance (skipping the second screen) by inputting the instance link rather than a registry link. This feature reinforces the idea that participation in any given registry is optional, not required. This is critical to avoiding vendor lock-in: discovery is treated as a service layered on top of the protocol, rather than a mandatory gateway. The protocol defines the minimum requirements and data shapes, while leaving higher-level concerns such as ranking, filtering, and governance to be implemented flexibly by individual registries.

\subsection{Mobile App}

\begin{figure}
    \centering
    \includegraphics[width=\linewidth]{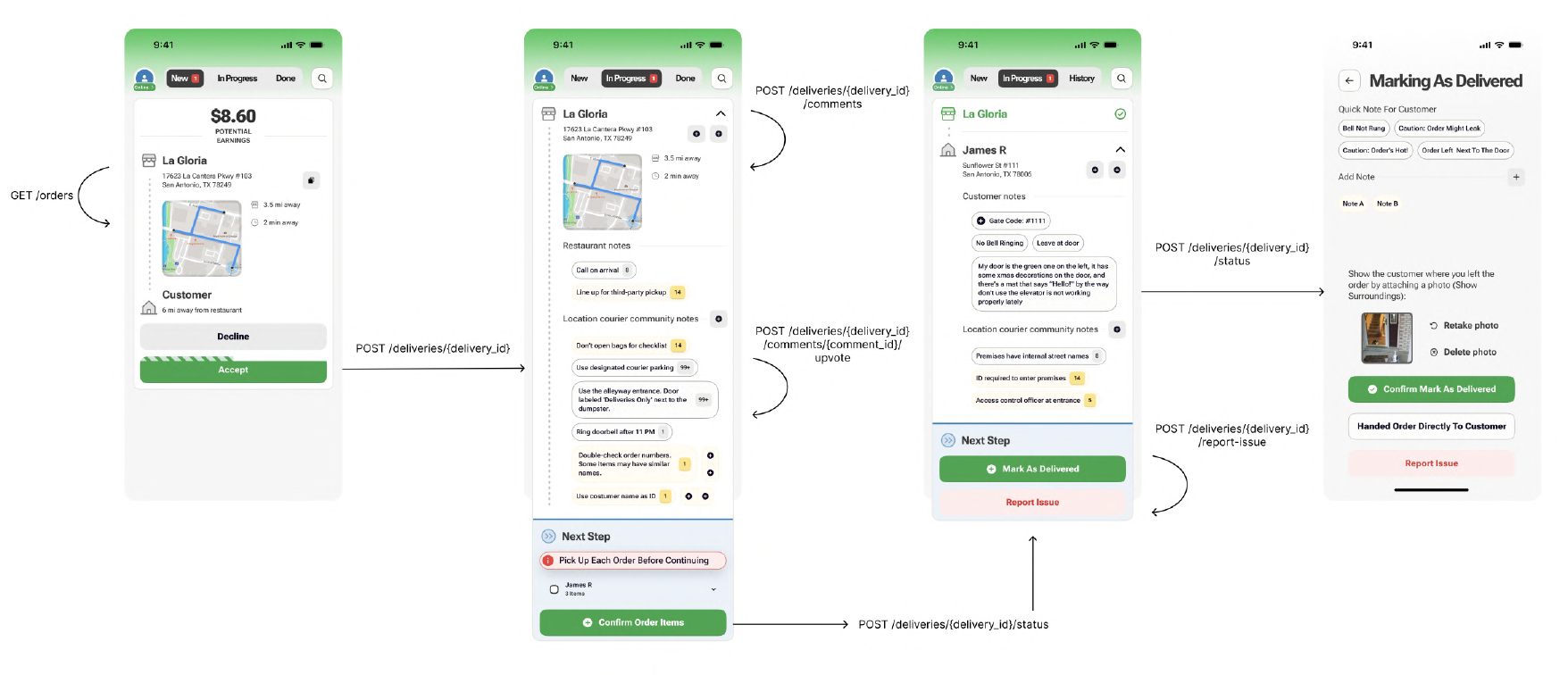}
    \caption{Screenshots of mobile app, showing how the delivery workflow looks for a courier using it.}
    \label{fig:workflow}
\end{figure}

We developed a mobile app that serves as a courier-facing client using React Native to ensure compatibility across both iOS and Android devices. The app connects to protocol-defined endpoints through dedicated UI components, enabling workers to fulfill deliveries with support from community notes and personalized preference inputs. The delivery fulfillment workflow through the app is illustrated in Figure~\ref{fig:workflow}.

\begin{figure}
    \centering
    \includegraphics[width=0.8\linewidth]{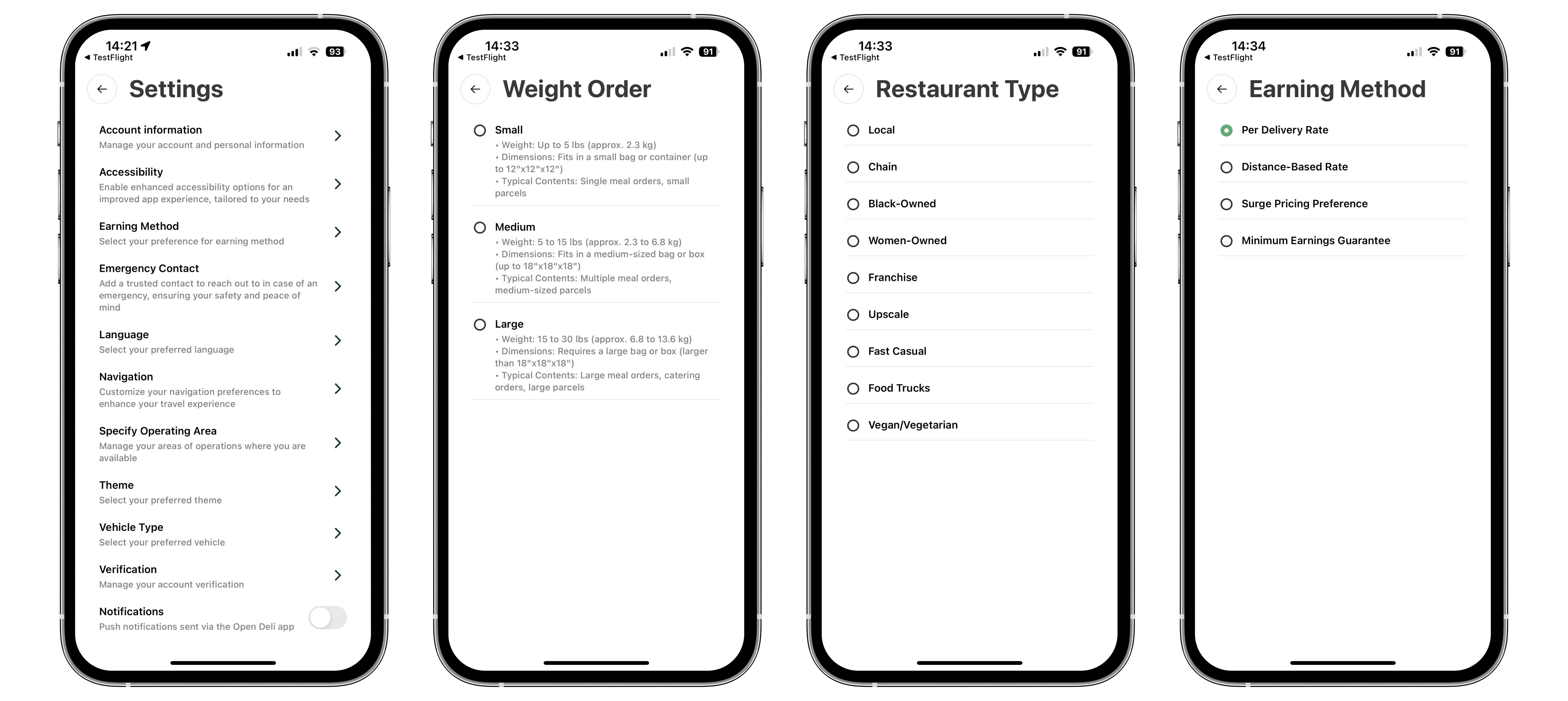}
    \caption{Screenshots of preference input interface in the mobile app client.}
    \label{fig:preference}
\end{figure}

Additionally, we built several settings pages where workers can input their preferences, as supported by the protocol (see Figure~\ref{fig:preference}). It is worth noting that the preference-input endpoints are more flexible than what is currently shown in the screenshots. Future work can expand the granularity and range of parameters supported, ensuring the system can better accommodate diverse needs and working styles.

\subsection{Backend Server Application and the Instance Admin Interface}
The backend server application implements the aforementioned endpoints to ensure the system is fully functional. A corresponding administrator interface provides a graphical user interface and visualization for non-experts to manage daily operations such as monitoring task statuses and editing instance settings at the instance level. The server application is built using the NestJS framework, with Prisma providing object-relational mapping for Node.js and TypeScript, and PostgreSQL as the underlying database. Passport handles authentication, while Swagger UI is used to visualize and document the API. For testing and deployment, the backend employs Jest and Docker, respectively. We implemented three example task-courier assignment algorithms: one that assigns tasks to the nearest available courier, another that prioritizes the most senior courier, and a third that assigns tasks to a specified courier, supporting system testing and enabling human intervention in the automation process. This implementation will be tested with real-life delivery workers and refined based on feedback soon. Our implementation is flexible and allows instance operators to deploy new collective decision-making models that incorporate diverse worker preferences, allowing matching algorithms to reflect the unique values of each instance.

In addition, we developed a user interface that enables instance administrators to configure instance-level settings and details as well as operational strategies, such as the algorithm used for courier-task matching, and tools for managing courier profiles and compensation. Administrators can register with any registries they have links for, view which registries they’re currently registered in, and unregister at any time.

\subsection{Future Implementations}
The protocol aims to encourage flexibility in how and by whom different components of the network are implemented. For example, the protocol does not foreclose the possibility of developing additional technical features that benefit a courier instance. A courier instance may want to design, develop, and vote on different task-assignment algorithms based on the priorities of its couriers. A mobile app it develops that compatible with the protocol may included additional voting features, as well as allow them to dynamically adopt the algorithms at different times as needed. 

The protocol also seeks to support interoperability, so that workers can move across points in the network fluidly. For example, a single mobile app can allow a courier to join and find work through multiple courier instances, which are possibly drawn from a variety of registries as well. This allows instances and couriers to adapt to diverse market needs and breaks the constraints in the current platform economy, where infrastructure and data are not interoperable. 


\end{document}